\documentclass[12pt]{article}

\usepackage{indentfirst}
\usepackage{amsmath}
\usepackage{graphicx}
\usepackage{float}
\usepackage{caption}
\usepackage{subcaption}
\usepackage{times}
\numberwithin{equation}{section}

\usepackage[toc,page]{appendix}

\usepackage{amssymb}
\usepackage{amsbsy}
\usepackage{amsmath}
\usepackage{mathtools}

\def\bit{\begin{itemize}}
\def\eit{\end{itemize}}
\def\beq{\begin{equation}}
\def\eeq{\end{equation}}
\def\bec{\begin{center}}
\def\eec{\end{center}}
\def\btable{\begin{tabular}}
\def\etable{\end{tabular}}
\def\beqr{\begin{eqnarray}}
\def\eeqr{\end{eqnarray}}
\def\beqrs{\begin{eqnarray*}}
\def\eeqrs{\end{eqnarray*}}
\def\benu{\begin{enumerate}}
\def\eenu{\end{enumerate}}
\def\noi{\noindent}
\def\half{\frac{1}{2}}

\def\del{\partial}

\def\btab{\begin{tabbing}}
\def\etab{\end{tabbing}}

\def\bit{\begin{itemize}}
\def\eit{\end{itemize}}

\def\Rarw{\Rightarrow}
\def\bfig{\begin{figure}}
\def\efig{\end{figure}}

\def\gm{\gamma}

\def\eps{\epsilon}

\def\bt{\beta}

\def\bfig{\begin{figure}}
\def\efig{\end{figure}}
\def\gm{\gamma}

\def\eps{\epsilon}

\def\bt{\beta}
\def\eps{\epsilon}

\def\Dl{\Delta}
\def\sg{\sigma}

\def\fr{\frac}
\def\non{\nonumber}

\usepackage{amsmath,amssymb}

\begin{document}
\title{Luminosity and beam-beam tune shifts with crossing angle and hourglass effects in e$^+$-e$^-$
colliders}

\author{Tanaji Sen \\ Accelerator Division, Fermilab \\ Batavia, IL 60510 }
\date{}

\maketitle

\begin{abstract}
  We develop theoretical expressions for the luminosity and beam-beam tune shifts in the presence of both a crossing angle
  and hourglass effects for the present and next generation of symmetric e$^+$-e$^-$ colliders. The theory is applied
  to the design of the Fermilab site filler Higgs factory and to the FCC-ee collider.
  \end{abstract}

\section{Introduction}

Introducing a crossing angle reduces both the luminosity and beam-beam tune shifts if no optics changes are
made. The lowered beam-beam tune shifts however allow the possibility of further reducing the beta functions
at the IP to increase the luminosity while keeping the beam-beam tune shifts within allowed limits. Specifically,
in an e$^+$-e$^-$ collider where $\bt_y^* \ll \bt_x^*$,  a crossing angle in the horizontal plane
allows a  scheme where $\bt_x^*$ is reduced sufficiently to increase the luminosity beyond values
without a crossing angle. This has been investigated in recent designs of colliders such as Super KEKB \cite{SupKEK},
FCC-ee \cite{FCC_2019} etc. We include  both the crossing angle and the
hourglass effects on the luminosity and the beam-beam tune shifts. Analytic expressions for the combined effects
do not appear to be available in the literature; instead they are approximated as acting independently.
The only assumption in our treatment below is that of symmetric interaction region optics for the electrons and
positrons so that the bunch sizes in all three dimensions are the same in both beams. After developing exact  general
expressions, we consider several limiting cases and show that they reduce to known forms where applicable.
The purpose of this paper is to find appropriate combinations of the crossing angle and $(\bt_x^*, \bt_y^*)$ which
maximize the luminosity while restricting the beam-beam tune shifts to tolerable values. 

We first apply the theory  to the very preliminary design of a Higgs factory at Fermilab, called a site filler,
and find parameters that increase the luminosity with  a non-zero crossing angle. Next we apply the theory to
the FCC-ee collider whose design is considerably more mature. We find that the luminosity in this design can also
be increased with changes in the crossing angle and $\bt_x^*$.

\section{Luminosity change with a crossing angle and hourglass effect}

The relativistically invariant luminosity per bunch and unit time is \cite{Handb}
\beq
    {\cal L}= K f_{rev} N_+ N_- \int_{-\infty}^{infty} \int_{-\infty}^{infty} \int_{-\infty}^{infty} \int_{-\infty}^{infty}
    ds\; dt\; dx\; dy \;     \rho_-(x y, s - ct) \rho_+(x, y, s - ct)
    \eeq
    ($N_-, \rho_-$), ($N_+, \rho_+$) are the (bunch intensities, three dimensional densities) of the electrons and positrons
    respectively and     $K$ is a kinematic factor 
    \beq
    K =  \sqrt{ ({\bf v_+} - {\bf v_-})^2 -  \fr{(\bf v_+ \times v_-)^2}{c^2} }
    \eeq
    We assume that the electrons move along the positive $s$ direction and the positrons in the opposite direction.
    When the beams cross in the horizontal plane at a full angle of $\theta_C$, 
the coordinates in the two beam frames are
    \beqr
    x_- & = &   C_C x  -  S_C s, \;\;\;  s_- =  C_C s + S_C x   \non \\
    x_+ & = & -  C_C x  -  S_C s, \;\;\;  s_+ = - C_C s + S_C x
    \eeqr
  where $(x, s)$ are the coordinates in the laboratory frame, $C_C= \cos(\theta_C/2), S_C = \sin(\theta_C/2) $. The transverse velocities are orders  of magnitude smaller
    than the longitudinal velocity ($\simeq c$), so the only velocity components are the projections of the longitudinal
    velocity, 
    \beqrs
        {\bf v_-} & \equiv &   \fr{d}{dt}(x_- \hat{x} + 0 \hat{y} +  s_-  \hat{z}) =
        (-S_C\hat{x} +  0\hat{y} + C_C\hat{z})c     \\ 
        {\bf v_+} & \equiv &   \fr{d}{dt}(x_+\hat{x} + 0\hat{y} + s_+\hat{z}) = (- S_C\hat{x} + 0\hat{y}
        - C_C\hat{z} )c   \\ 
          ({\bf v_-} - {\bf v_+})^2 & = &  ( 0\hat{x} + 0\hat{y} +2 C_C\hat{z} )^2 c^2  = 4 C_C^2 c^2  \\
          (\bf v_+ \times v_-) & = & (0\hat{x} +  2 S_C C_C \hat{y} + 0\hat{z}) c^2 
  \eeqrs
where $\hat{x}, \hat{y}, \hat{z}$ are unit vectors.  Hence the kinematic factor is
          \beq
          K = c \sqrt{4 C_C^2 - 4 S_C^2 C_C^2} = 2 c C_C^2
          \eeq 
    The normalized density of the electron and positron bunches are
\beqr
    \rho_-  \!\!\! &  = &  \!\!\! \fr{1}{(2\pi)^{3/2} \sg_{x-} \sg_{y -} \sg_{s -}}
      \exp\left[ - \fr{( C_C x  -  S_C s)^2}{2\sg_{x -}^2} - \fr{y^2}{2\sg_{y-}^2} 
 - \fr{( C_C s + S_C x - ct)^2}{2 \sg_{s -}^2}\right]   \\
          \rho_+ \!\!\! \!\!\! &  = &  \!\!\!\!\!\! \fr{1}{(2\pi)^{3/2} \sg_{x+} \sg_{y +} \sg_{s +}}
   \exp\left[ - \fr{( -C_C x  -  S_C s)^2}{2\sg_{x +}^2} - \fr{y^2}{2\sg_{y+}^2}
 - \fr{( S_C x  - C_C s  - ct)^2}{2 \sg_{s +}^2} \right]   
      \eeqr
      We assume that the bunches are symmetric at the IP so that the beam sizes in all 3 dimensions are matched, i.e.
      \beq
      \sg_{x +}^* =  \sg_{x -}^*, \;\;\;     \sg_{y +}^* =  \sg_{y -}^*,  \;\;\;      \sg_{s +} =  \sg_{s -} = \sg_s 
      \eeq
      This assumption is true for equal high energy colliders that we consider but could be dropped
      to consider the  more general case of asymmetric colliders. We ignore perturbative effects such as a non-zero dispersion or
       transverse offsets at       the IP. 
      
      The beta functions depend only on the longitudinal coordinate $s$ in the lab frame, they do not depend on  the  coordinates in the beam frames. Thus
      \beqr
      \sg_{x -}(s) &  = & \sg_{x +}(s) = \sg_x(s)  = \sqrt{\eps_x(\bt_x^*+  \fr{s^2}{\bt_x^*}) } =
      \sg_x^*\sqrt{1 + \fr{s^2}{\bt_x^{*, 2}} }  \\
      \sg_{y -}(s)  &  = &  \sg_{y +}(s) = \sg_y(s)  = \sg_y^*\sqrt{1 + \fr{s^2}{\bt_y^{*, 2}} }
      \eeqr 

 Putting all the factors for the luminosity
 \beqrs
     {\cal L} &= &  \fr{2c f_{rev} N_+ N_- C_C^2}{ (2\pi)^{3}}\int_{-\infty}^{\infty} \int_{-\infty}^{\infty} \int_{-\infty}^{\infty}
     \int_{-\infty}^{\infty}ds \; dt  \; dx  \; dy \; \fr{1}{ \sg_{x}(s)^2 \sg_{y}(s)^2 \sg_{s}^2}  
     \exp[ - \fr{( C_C x  -  S_C s)^2}{2\sg_{x }^2}]  \\
&  &     \exp[- \fr{y^2}{2\sg_{y}^2}]   \exp[ - \fr{( C_C s + S_C x  - ct)^2}{2 \sg_{s}^2}]  
     \exp[ - \fr{( -C_C x  -  S_C s)^2}{2\sg_{x }^2}]\exp[- \fr{y^2}{2\sg_{y}^2}] \\
&  &      \exp[ - \fr{( C_C s - S_C x  + ct)^2}{2 \sg_{s}^2}]   
\eeqrs 
The integrations over $t, y, x$ are straightforward.   The last integration over $s$ is
 \beqrs
 &   &    \int_{-\infty}^{\infty} \fr{ ds}{\sg_s} \;  
 \fr{1}{\sqrt{(1 +\fr{s^2}{\bt_x^{* 2}})(1 +\fr{s^2}{\bt_x^{* 2}})}}
 \exp[- \fr{  C_C^2 s^2 }{\sg_{s}^2} - \fr{S_C^2 s^2}{\sg_x^2}]  \\
 & = &   \int_{-\infty}^{\infty} du \; \exp[ - C_C u^2( 1 + T_C^2 \fr{\sg_s^2/\sg_x^{*2}}{1 + u^2/( u_x^2)}) ]
 \fr{1}{\sqrt{ (1 + \fr{u^2}{ u_x^2}) (1 + \fr{u^2}{ u_y^2})}}
\eeqrs
 where we defined
 \beq
 u = \fr{s}{\sg_s}, \;\;\; u_x = \fr{\bt_x^*}{\sg_s}\;\;\; u_y = \fr{\bt_y^*}{\sg_s} \;\;\;  T_C = \tan(\theta_C/2)
 \eeq
 Thus the general expression for the luminosity per bunch is
 \beqr
     {\cal L}   & = &  \fr{f_{rev}  N_+ N_- C_C}{4 \pi^{3/2} \sg_x^* \sg_y^* }
       \int_{-\infty}^{\infty} du \; \exp[ - C_C^2 u^2( 1 + T_C^2 \fr{\sg_s^2/\sg_x^{*2}}{1 + u^2/( u_x^2)}) ]
       \fr{1}{\sqrt{ (1 + \fr{u^2}{ u_x^2}) (1 + \fr{u^2}{ u_y^2})}} \non \\
       \label{eq: hg_cross}
  \eeqr
  and the general correction factor is
   \beqr
 R_L   \!\!\! & \equiv & \fr{\cal L}{\cal L_0} \non \\
   & = &   \!\!\!  \fr{C_C}{\sqrt{\pi}}
       \int_{-\infty}^{\infty} du \; \exp[ - \cos^2(\theta_C/2) u^2( 1 + \tan^2(\theta_C/2)  \fr{\sg_s^2/\sg_x^{*2}}{1 + u^2/( u_x^2)}) ]
       \fr{1}{\sqrt{ (1 + \fr{u^2}{ u_x^2}) (1 + \fr{u^2}{ u_y^2})}} \non \\
       \label{eq: RL_gen}
 \eeqr

\noi {\bf  Limiting cases}
  \benu
\item \underline{No crossing angle or hourglass } 

  Setting $C_C= 1, T_C=0$ and  $u_x, u_y \to \infty $, we have
  \beq
      {\cal L}  =  \fr{ f_{rev} N_+ N_- }{4 \pi^{3/2} \sg_x^* \sg_y^* }        \int_{-\infty}^{\infty} du \; \exp[ - u^2]
      = \fr{ f_{rev} N_+ N_- }{4 \pi \sg_x^* \sg_y^* }
      \eeq
      the standard expression for the nominal luminosity.

    \item \underline{Only the hourglass effect, no crossing angle}
\beq
     {\cal L}  =  \fr{f_{rev}  N_+ N_- }{4 \pi^{3/2} \sg_x^* \sg_y^* }
       \int_{-\infty}^{\infty} du \; \exp[ -  u^2 ]  \fr{1}{\sqrt{ (1 + \fr{u^2}{u_x^2}) (1 + \fr{u^2}{u_y^2})}}
  \eeq
  Hence the luminosity correction factor is
  \beq
  R_L \equiv \fr{\cal L}{\cal L_0} =  \fr{1}{\sqrt{\pi}}
       \int_{-\infty}^{\infty} du \; \exp[ -  u^2 ]
 \fr{1}{\sqrt{ (1 + \fr{u^2}{u_x^2}) (1 + \fr{u^2}{u_y^2})}}
  \eeq
  which agrees with the  expression Eq.(2.7) in \cite{Furman}.

  \item \underline{Only the crossing angle, no hourglass effect}

  In the limit that $\bt_x^*, \bt_y^* \gg \sg_z^*$, there is very little variation in the transverse sizes across
  the bunch length. In this limit $(t_x^2, t_y^2 \gg 1$. The dominant contributions to the integral come from the regions
  close to $t=0$ because of the exponential factor. In this limit we can assume $(t/t_x)^2, (t/t_y)^2 \ll 1$ and the
  luminosity is
 \beqr
     {\cal L}   & = &  \fr{f_{rev}  N_+ N_-  C_C}{4 \pi^{3/2} \sg_x^* \sg_y^*}
       \int_{-\infty}^{\infty} du \; \exp[ - C_C^2 u^2( 1 + T_C^2 \fr{\sg_s^2}{\sg_x^{*2}}) ]  \non \\ 
       & = &  \fr{f_{rev}  N_+ N_- C_C }{4 \pi^{3/2} \sg_x^* \sg_y^* }
       \fr{\sqrt{\pi}}{C_C\sqrt{( 1 + T_C^2 \fr{\sg_s^2}{\sg_x^{*2}})}}
  \eeqr
  Hence the luminosity correction factor is
  \beq
  R_L =        \fr{1}{\sqrt{( 1 + T_C^2 \fr{\sg_s^2}{\sg_x^{*2}})}}
\eeq
This is the standard correction factor for the crossing angle. 

  
\item \underline{Flat bunch}

  In this limit,  $\sg_x^*  \gg \sg_y^*$. Since the equilibrium emittances obey   $\eps_x \gg \eps_y$,  this
  is easily satisfied if $\bt_x^* \gg \bt_y^*$, which typically is the case. 
  In this limit we drop  all terms with $u_x$ from Eq.(\ref{eq: hg_cross})
  \beqr
     {\cal L}   & = &  \fr{f_{rev}  N_+ N_- C_C}{4 \pi^{3/2} \sg_x^* \sg_y^* }
 \int_{-\infty}^{\infty} du \; \exp[ - C_C^2 u^2( 1 + T_C^2 \fr{\sg_s^2}{\sg_x^{*2}}) ]
 \fr{1}{\sqrt{ (1 + \fr{u^2}{ u_y^2})}}  
  \eeqr
The luminosity correction factor is 
  \beqr
R_{flat} & = &
  \fr{C_C}{\sqrt{\pi}} \int_{-\infty}^{\infty}du \; \exp[ - C_C^2 u^2( 1 + T_C^2 \fr{\sg_s^2}{\sg_x^{*2}}) ]
 \fr{1}{\sqrt{ (1 + \fr{u^2}{ u_y^2})}}   \\ 
    & = & \fr{C_C}{\sqrt{\pi}} u_Y \exp[\half b^2 u_y^2] K_0(\half b^2 u_y^2)    \label{eq: lumi_flat}
  \eeqr
  where
  \beq
  b^2 = C_C^2[1 +   T_C^2 \fr{ \sg_z^2}{\sg_x^{*2} }] \ge 0 
  \eeq
  and $K_0$ is a Bessel function. There is a similar expression Eq. (2)   in \cite{Hirata}. 

  In the absence  of a crossing angle so that $C_C = 1, T_C = 0$, we have $b= 1$
  and Eq, \ref{eq: lumi_flat} is the same as Eq. (2.12) in Furman.

\eenu

  \section{Beam-beam tune shifts}
  
The beam-beam potential for electrons interacting with a positron bunch which has a longitudinally Gaussian density is
\beqr
U(x,y) & = &  \frac{N_+ r_e}{\gm_e} \int \fr{\exp[-s^2/(2\sg_s^2)]}{\sqrt{2\pi}\sg_s}     \; ds
\int_0^\infty  dq\; \frac{1}{(2\sg_x(s)^2+q)^{1/2}(2\sg_y(s)^2+q)^{1/2}} \non  \\
&  &          \left\{1- \exp[-\frac{x^2}{2\sg_x(s)^2+q} - \frac{y^2}{2\sg_y(s)^2+q}] \right\} 
\eeqr
where the parameters are those of the positron bunch. 
The potential for electrons interacting with positrons at a full crossing angle $\theta_C$ can be
found by replacing
\beq
x \to   C_C x  -  S_C s, \;\;\;  s \to  C_C s + S_C x  \label{eq: rot_x}
\eeq

The amplitude dependent beam-beam tune shifts can be obtained from the second derivatives of the potential as 
\beq
\Dl \nu_x(x, y) = -\frac{\bt_x^* N_+ r_e }{4 \pi \gm_e} \fr{\del^2 U}{\del x^2}, \;\;\;
\Dl \nu_y(x, y) = -\frac{\bt_y^* N_+ r_e }{4 \pi \gm_e} \fr{\del^2 U}{\del y^2}
\eeq
The beam-beam tune shift parameters are the values of the tune shifts at the origin, i.e.
\beq
\xi_x = \Dl \nu_x(0, 0), \;\;\; \xi_y = \Dl \nu_y(0, 0)
\eeq

Substituting the rotated forms in Eq.(\ref{eq: rot_x}) into the potential, taking the derivatives, evaluating the
terms at the origin and leads to
\beqr
\xi_x & = & \fr{\bt_x^* N_+ r_e }{2 \pi \gm_e} \int_{0}^{\infty}\fr{ds}{\sqrt{2\pi} \sg_s} \int_0^{\infty} dq\; 
2 \fr{\exp[- \fr{s^2 C_C^2}{(2 \sg_s^2)} - \fr{(s^2 S_C^2)}{(q+2 \sg_x^2(s))} ]}
{\sqrt{(q+2 \sg_x^2(s))^3 (q+2 \sg_y^2(s))}}  \non \\
& & \times  C_C^2   \left\{ 1  +2 s^2  S_C^2[ \fr{1}{\sg_s^2}  - \fr{1}{q+2 \sg_x^2(s)} ] \right\} \non  \\
 &   &   - \fr{\exp[- \fr{s^2 C_C^2}{2 \sg_s^2}]\left(1 - \exp[- \fr{s^2 S_C^2}{q+2 \sg_x^2(s)}] \right)}
 {\sg_s^2 \sqrt{(q+2 \sg_x^2(s)) (q+2 \sg_y^2(s))}}S_C^2 \left[ 1 - \fr{s^2 C_C^2 }{\sg_s^2} \right]     \label{eq: xix_1}  \\
 \xi_y  & = & \frac{\bt_y^* N_+ r_e }{2 \pi \gm_e }\int_{-\infty}^{\infty} \fr{ds}{\sqrt{2\pi}\sg_s}
 \int_0^{\infty} dq    \frac{2 \exp \left(-\frac{t^2 S_c^2 \sg_s^2}{q+2 \sg_x^2}-
  \frac{t^2 C_C^2 }{2 }\right)}{\left(q+2   \sg_y^2\right)
   \sqrt{\left(q+2 \sg_x^2\right) \left(q+2    \sg_y^2\right)}} \label{eq: xiy_1}
   \eeqr

   The different integrations  over $s$ cannot all be done analytically, so the 2D integrations have to be done
   numerically. Transform to dimensionless variables $(u,t)$ and define other dimensionless variables
\beqr
t & = &  \fr{s}{\sg_s}, \;\;\;   t_x = \fr{\bt_x*}{\sg_s}, \;\;\;   t_y = \fr{\bt_y*}{\sg_s} \\
u & = &  \fr{2\sg_x^{*, 2}}{q + 2\sg_x^{*, 2}} \;\;\; \Rarw   q = 2 \sg_x^{*,2}(\fr{1}{u } - 1); \;\;\;
0\le u \le 1  \\ 
& \Rarw  & \sg_x(t)=\sg_x^*\sqrt{1 +  \fr{s^2}{\bt_x^{*, 2}}} = \sg_x^*\sqrt{1 +  \fr{t^2}{t_x^2}} \\
&  \Rarw  &   \sg_y(t) = \sg_y^*\sqrt{1 +  \fr{s^2}{\bt_y^{*, 2}}} = \sg_y^*\sqrt{1 +  \fr{t^2}{t_y^2}} \\
r_{yx} &  =  &  \fr{\sg_y^{* 2}}{\sg_x^{* 2}}, \;\;\; r_{sx} = \fr{\sg_s^2}{2 \sg_x^{* 2}}
\eeqr
The Jacobian of the transformation is $J(q, s; u, t)   =   \fr{2\sg_x^{*,2}\sg_s}{u^2} $. 
Carrying out the transformation leads to the equations for the general case
\beqr
  \xi_x & = &  \frac{\bt_x^* N_p r_e }{2 \pi \gm_e}  \left\{\fr{ C_C^2}{(\sg_x^{*, 2})}
   \int_{0}^{\infty}  \fr{dt}{\sqrt{2\pi}}  \;   \exp[- \fr{ C_C^2 t^2}{2 }] 
\int_0^{1} du \; \exp[ -  \fr{( r_{sx} S_C^2 u t^2)}{1 +  u (t/t_x)^2}]  \right.  \non \\
    &  &  \times   \left( 1  + 2  S_C^2 t^2[ 1  -  r_{sx}\fr{u}{1 +  u (t/t_x)^2} ] \right) 
  \left[ \fr{1}{(1 +  u (t/t_x)^2)^3 (1 + [ r_{yx}(1 +  (t/t_y)^2) - 1] u )} \right]^{1/2}   \non  \\
    &   &  - \fr{S_C^2}{\sg_s^2 } \int_{0}^{\infty}  \fr{dt}{\sqrt{2\pi}}  \; 
\exp[- \fr{ C_C^2 t^2}{2}]     \left[ 1 -  C_C^2 t^2 \right]  
 \int_0^{1} \fr{du }{u}  \; \left(1 - \exp[-   \fr{ r_{sx}  S_C^2 u t^2}{1 +  u (t/t_y)^2}] \right)  \non \\ 
 &  &  \left. \times \left[ \fr{1}{(1 +   u (t/t_x)^2) (1 + [ r_{yx}(1 +  (t/t_y)^2) - 1] u)} \right]^{1/2}
 \right\}   \label{eq: xix_all} \\
\xi_y      & = &  \frac{\bt_y^* N_p r_e }{ \pi \gm_e } \fr{1}{(2\sg_x^{*,2})}
   \int_{0}^{\infty} \fr{dt}{\sqrt{2\pi}} \; \exp[- \fr{ C_C^2 t^2}{2}]  
   \int_0^{1} du \; \exp \left[-   \fr{r_{sx} S_C^2 u t^2}{(1 +  u (t/t_x)^2)}\right] \non \\
   &  &  \times \left[\fr{1}{(1 +   u  (t/t_x)^2)(1 + [ r_{yx}(1 +   (t/t_y)^2) - 1] u)^3}\right]^{1/2} 
 \label{eq: xiy_all}
\eeqr
These expressions involve a finite range of integration (from $0 \to 1$) over $u$ compared to the infinite
range over $q$ in Eq.(\ref{eq: xix_1}) and (\ref{eq: xiy_1}). In those equations, the convergence rate is poor
and evaluate  slowly while  the double integrals evaluate quickly in the set Eq.(\ref{eq: xix_all}) and
(\ref{eq: xiy_all}) above. We note that the second set of terms in Eq.(\ref{eq: xix_all}) proportional to
$\fr{S_C^2}{\sg_s^2 }$ are typically very small in the collider applications considered below and can be
ignored.

\vspace{2 em}

\noi {\bf  Limiting cases}
\benu
\item \underline{No crossing angle or hourglass effect}

  In this case $S_C = 0, C_C = 1$,  so
\beqrs
\xi_x & = &   \frac{\bt_y^* N_+ r_e }{2 \pi \gm_e}  t_x  \fr{2}{(2\sg_x^{*, 2})}
\int_{0}^{\infty}  \fr{dt}{\sqrt{2\pi}}  \exp[- \fr{t_x^2  t^2}{2 }] 
\int_0^{1} du \; \left[ \fr{1}{(1 + (r_{yx}-1) u )} \right]^{1/2}    \\
\xi_y & = &   \frac{\bt_y^* N_+ r_e }{\pi \gm_e } \fr{t_x}{(2\sg_x^{*,2})}
\int_{0}^{\infty} \fr{dt}{\sqrt{2\pi}}\exp[- \fr{t_x^2 t^2}{2}]
\int_0^{1} du   \left[\fr{1}{(1 + (r_{yx}-1)u)^3} \right]^{1/2}
\eeqrs

Using Eqs.(\ref{eq: int_xix_nom}) and (\ref{eq: int_xiy_nom}) in the Appendix, we find 
  \beqr
  \xi_x & = &    \frac{\bt_x^* N_+ r_e }{2 \pi \gm_e  }   \fr{1}{ \sg_x^{*}(\sg_y^{*} + \sg_x^{*})}
  \label{eq: xix_nom} \\ 
  \xi_y & = &    \frac{\bt_y^* N_+ r_e }{2 \pi \gm_e } \fr{1}{\sg_y^{*}(\sg_y^{*}+ \sg_x^{*})}
  \label{eq: xiy_nom}
  \eeqr
These are the standard expressions for the tune shifts without a crossing angle or hourglass effects.

\item \underline{Only a crossing angle }

We  use the expressions Eq.(\ref{eq: xix_all}) and (\ref{eq: xiy_all}), let $t_x, t_y \to \infty$ and do the integration over $t$ first.  Define shorthand variables
 \beq
   a^2 =  r_{sx}  S_C^2 , \;\;\;  c^2 =  \fr{1}{2}  C_C^2
   \eeq
  We use the integration results in Eqs. (\ref{eq: xi_cr1}) - (\ref{eq: xi_cr3}) to obtain
  \beqr
  \xi_x  & = &  \frac{\bt_x^* N_+ r_e }{2 \pi \gm_e}  \fr{1}{4\sqrt{2}} \int_0^{1} du \;
  \left[\fr{1}{(1 + [ r_{yx} - 1] u)} \right]^{1/2}
  \left\{\fr{2  C_C^2}{(2\sg_x^{*, 2})}   \fr{2c^2 + 2 t_x^2  S_C^2 }{(c^2 + a^2 u)^{3/2}}  \right. \non \\
&  &  \left. + \fr{S_C^2}{\sg_s^2 } \fr{1 }{u}   \fr{2 a^2 u }{(c^2 + a^2 u)^{3/2}} \right\}\non\\
  & = &   \frac{\bt_x^* N_+ r_e }{2 \pi \gm_e} \fr{1 }{C_C^3}
\fr{1}{ [\sg_y^{*} + \sqrt{\sg_x^{* 2} + T_C^2 \sg_s^{ 2}}]\sqrt{\sg_x^{* 2} + T_C^2 \sg_s^{ 2}}} \\
\xi_y & = &  \fr{\bt_y^* N_+ r_e }{2 \pi \gm_e }\fr{1}{C_C} \int_0^{1)} du\fr{1}{\sqrt{(2(\sg_y^2 - \sg_x^2)u + 1)^3[ 1 + 2 (T_C^2 \sg_s^2)u/(2 \sg_x^2) ]}} \non \\
& = & \fr{\bt_y^* N_+ r_e }{2 \pi \gm_e } \fr{1}{C_C}\fr{1}{\sg_y(\sg_y + \sqrt{\sg_x^2 + T_C^2\sg_s^2})}
 \eeqr
 These expressions can be obtained from the equations (\ref{eq: xix_nom}), (\ref{eq: xiy_nom}) by replacing the
 transverse beam size $\sg_x$ in the crossing plane by the effective beam size
 $\sqrt{\sg_x^2 + T_C^2\sg_s^2})$
 and including the factors $1/C_C^3$ in $\xi_x$, $1/C_C$ in $\xi_y$ which are $\sim 1$ for typical crossing angles.

\item \underline{Only the hourglass, no crossing angle }

Setting $C_C= 1,  S_C=0,   = 1$ in the general forms Eq.(\ref{eq: xix_all}), and (\ref{eq: xiy_all})
\beqrs
  \xi_x  \!\!\! & = &  \!\!\! \frac{\bt_x^* N_+ r_e }{2 \pi \gm_e}  \left\{\fr{2 }{(2\sg_x^{*, 2})}
  \int_{0}^{\infty}  \fr{dt}{\sqrt{2\pi}}   \exp[- \fr{ t^2}{2 }]  \right.  \\
& & \left. \times  \int_0^{1} du \;   \left[ \fr{1}{(1 +u t^2/t_x^2)^3 (1 + [ r_{yx}(1 +  t^2/t_y^2) - 1] u )} \right]^{1/2}     \right\} \\
\xi_y      & = &  \frac{\bt_y^* N_p r_e }{ \pi \gm_e } \fr{1}{(2\sg_x^{*,2})}
   \int_{0}^{\infty} \fr{dt}{\sqrt{2\pi}} \; \exp[- \fr{ t^2}{2}]   \\
& & \times
   \int_0^{1} du \; \left[\fr{1}{(1 +   u  (t/t_x)^2)(1 + [ r_{yx}(1 +   (t/t_y)^2) - 1] u)^3}\right]^{1/2} 
\eeqrs
Integrating over $u$ yields the expressions 
  \beqr
  \xi_x & = &   \frac{\bt_x^* N_+ r_e }{ \pi \gm_e \sg_x^{*, 2}} 
\int_{0}^{\infty}  \fr{dt}{\sqrt{2\pi}}  \exp[-t^2/2]
\fr{1}{\sqrt{(1 + \fr{t^2}{t_x^2} ) }
\left[  \sqrt{ 1 + \fr{t^2}{t_x^2}}+ \fr{\sg_y^*}{\sg_x^*}\sqrt{1 + \fr{t^2}{t_y^2}} \right]}     \\
   \xi_Y  & = &  \frac{\bt_y^* N_+ r_e }{2 \pi \gm_e \sg_y^{*, 2}} \int \fr{dt}{\sqrt{2\pi}} \exp[-t^2/2]
\fr{1}{\sqrt{(1 + \fr{t^2}{t_y^2} )}
\left[  \sqrt{ 1 + \fr{t^2}{t_y^2}}+ \fr{\sg_x^*}{\sg_y^*}\sqrt{1 + \fr{t^2}{t_x^2}} \right]}
\eeqr
These agree with Eq.(3.4) (evaluated at $s= 0$) in\cite{Furman}.

\item \underline{Flat beams} \\ 
  Here we consider the limit $t_x \to \infty$  in the general expressions in Eqs (\ref{eq: xix_all}), (\ref{eq: xiy_all}). Interchanging the integrations and we drop the second set of terms $\propto S_C^2$ in $\xi_x$ that are
negligibly small
  \beqr
  \xi_x & = &  \frac{\bt_x^* N_+ r_e }{2 \pi \gm_e}   \fr{2 C_C^2}{(2\sg_x^{*, 2})}
\int_0^{1} du \;    \int_{0}^{\infty}  \fr{dt}{\sqrt{2\pi}}    
\exp[ -   t^2(\half C_C^2 + r_{sx}  S_C^2 u )]   \non \\
    &  &  \times   \left( 1  + 2  S_C^2 t^2[ 1  -  r_{sx} u ] \right) 
\left[ \fr{1}{(1 + (r_{yx}- 1)u  +  r_{yx}u t^2 /t_y^2  )} \right]^{1/2}  \label{eq: xix_fl_1}\\
\xi_y      & = &  \frac{\bt_y^* N_+ r_e }{ \pi \gm_e } \fr{1}{(2\sg_x^{*,2})}
   \int_0^{1}   \int_{0}^{\infty} \fr{dt}{\sqrt{2\pi}} \exp[ - t^2(\half C_C^2 + r_{sx}  S_C^2 u )] \non \\
   &  &  \times \left[\fr{1}{(1 + (r_{yx}- 1)u  +  r_{yx}u t^2 /t_y^2  )} \right]^{3/2}  \label{eq: xiy_fl_1}
   \eeqr
The integrations over $t$ can be done and expressed in terms of confluent hypergeometric function $U$ and
the Bessel function $K_0$. Using integration results in the Appendix and identifying the coefficients in the
equations (\ref{eq: flat_1}) and (\ref{eq: flat_2})
\beqrs
a^2 & = &  \half C_C^2 + r_{sx}  S_C^2 u, \;\; ({\rm Note} \;\; u \ge 0),\;\;\; b = 2  S_C^2 [ 1  -  r_{sx} u ]  \\
c  & = & 1 + (r_{yx}- 1)u , \;\;\; d =  r_{yx}u /t_y^2  
\eeqrs
 leads to the expressions for the tune shifts in the flat beam limit as
  \beqr
  \xi_x & = &  \frac{\bt_x^* N_+ r_e }{2 \pi \gm_e}   \fr{2 C_C^2}{2\sg_x^{*, 2}\sqrt{2\pi}}
\int_0^{1} du \;  \fr{1}{4(\half C_C^2 + r_{sx}  S_C^2 u) \sqrt{r_{yx}u /t_y^2}}  \non \\
& & \left\{ (  C_C^2 + 2 r_{sx}  S_C^2 u) \exp[ {\rm arg}] K_0({\rm arg})  
+ \sqrt{\pi} 2  S_C^2 [ 1  -  r_{sx} u ] U(\half, 0, 2\; {\rm arg}))  \right\}   \non \\
&  &  \mbox{}   \label{eq: xix_fl_2}\\ 
\xi_y      & = &  \frac{\bt_y^* N_+ r_e }{ \pi \gm_e } \fr{1}{2\sqrt{2}\sg_x^{*,2}}
\int_0^{1} du \; \fr{1}{2(1 + (r_{yx}- 1)u ) \sqrt{r_{yx}u /t_y^2  }}U(1/2, 0, 2\; {\rm arg})
\non  \\
\mbox{}   \label{eq: xiy_fl_2}  \\
{\rm arg} & = & \fr{(\half C_C^2 + r_{sx}  S_C^2 u)(1 + (r_{yx}- 1)u)}{ 2 r_{yx}u /t_y^2 }]
   \eeqr
   This in principle, leaves the integration over $u$ to be evaluated numerically. While the integrations over
   the hypergeometric function $U$ in both $\xi_x, \xi_y$ converge rapidly for typical parameter values, the
   integration over the
   Bessel function $K_o$ in $\xi_x$ is poorly convergent. We found it more convenient to use the double
   integration in Eq.(\ref{eq: xix_fl_1}) to evaluate $\xi_x$ and the single integration in Eq.(\ref{eq: xiy_fl_2})
   to evaluate $\xi_y$.   In the following sections, we have used the exact expressions  for the luminosity and
   the beam-beam tune shifts for both the Fermi site filler and the FCC-ee but in both cases, the flat beam
   expressions are very good approximations.

\eenu

\section{Higgs Factory Site Filler}

\begin{table}[htb]
  \bec
  \btable{|c|c|} \hline
 Beam energy [GeV]  & 120 \\
  Circumference [km] &  16.0  \\
  Bunch intensity  &  8.34 $\times 10^{11}$ \\
  Number of bunches &  2  \\
  Emittance x [nm] / y [pm] & 0. /   \\
  $\bt_x^* / \bt_y^*$ [m]  & 0.2 / 0.001   \\
  $\sg_z$ [mm] &  2.9 \\
Nominal crossing angle [mrad] &  0 \\
  \hline
  \etable
  \caption{Parameters of the Fermilab site filler }
  \label{table: SF_param}
  \eec
  \end{table}
Table \ref{table: SF_param} shows the parameters in the very preliminary design of a Higgs factory based
at Fermilab \cite{Fermi_SF}. The bunch length in the table above is the equilibrium value with only
synchrotron radiation emitted in the arcs and does not include the beamsstrahlung effect that would increase
the bunch length. Hence, the hourglass effects for the site filler are a slight underestimate of the exact values. 
Table \ref{table: FSF_lumibb} shows the luminosity and the beam-beam tune shifts with and without the hourglass
effect. 
\begin{table}[htb]
  \bec
  \btable{|c|c|c|} \hline
  &    Without hourglass  & With hourglass   \\  \hline
  Luminosity  [cm$^{-2}$-s$^{-1}$] & 1.56 $\times 10^{34}$  &   9.5 $\times 10^{33}$        \\
  Beam-beam tune shift $\xi_x, / \xi_y$ &    0.00849 / 0.1197  & 0.00845 / 0.0614  \\
  \hline
  \etable
  \caption{Luminosity and beam-beam tune shifts in the Fermi site filler with the parameters shown in Table
    \ref{table: SF_param}. }
    \label{table: FSF_lumibb}
  \eec
  \end{table}

\bfig
\centering
\includegraphics[scale=0.4]{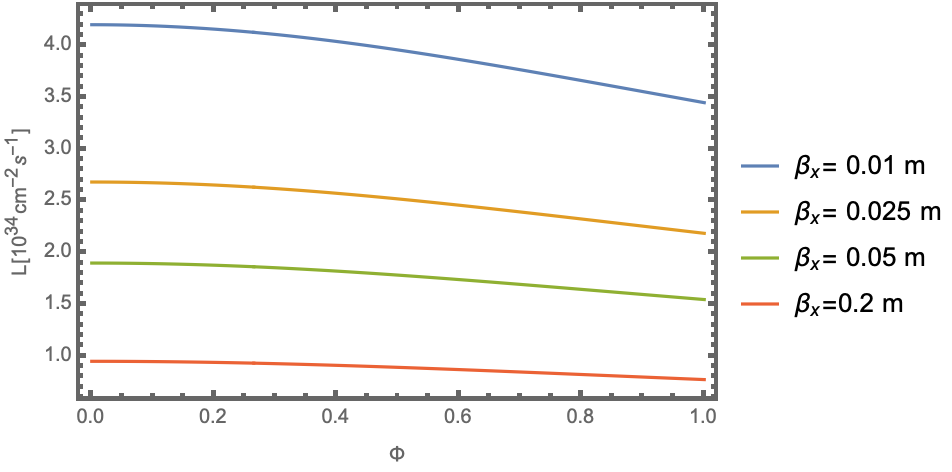}
\includegraphics[scale=0.65]{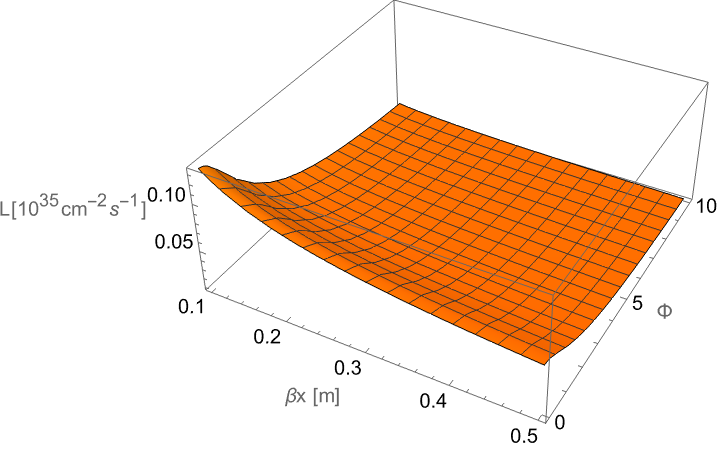}
\caption{(Left): Luminosity as a function of the Piwinski angle $\Phi$ for four values of $\bt_x*$. (Right):
  Luminosity as a function   of $\Phi$ and $\bt_x^*$.  $\bt_y^*$ is constant at 1mm in both figures. }
\label{fig: Lumi}
%
\centering
\includegraphics[scale=0.55]{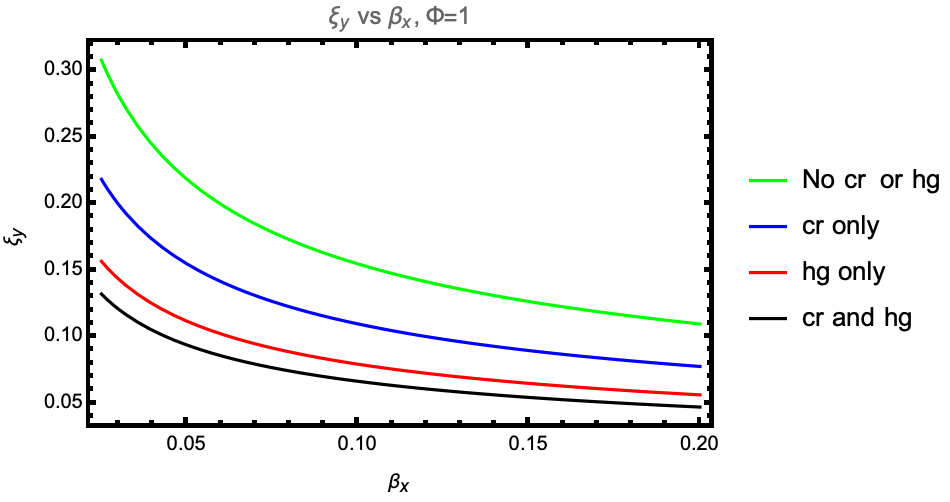}
\includegraphics[scale=0.55]{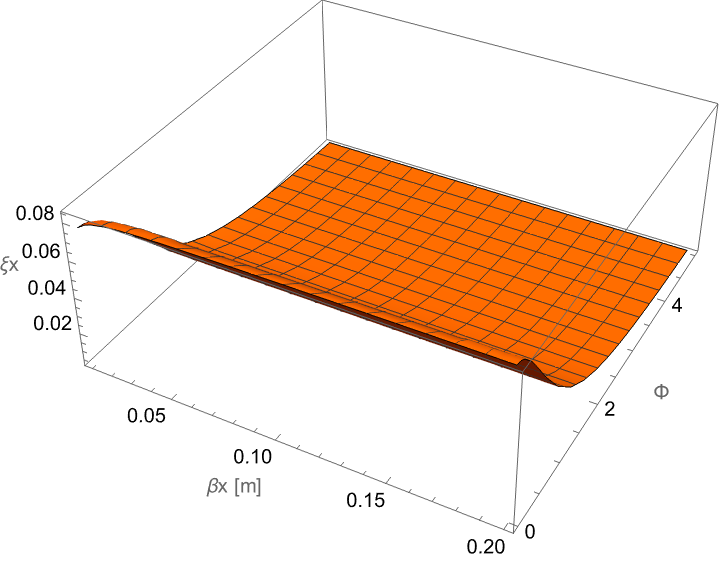}
\includegraphics[scale=0.55]{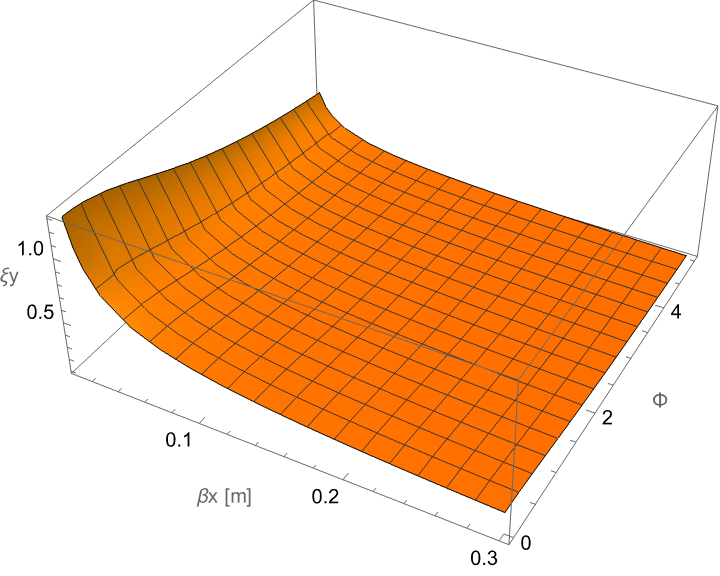}
\caption{Top: Vertical beam-beam tune shift $\xi_y$ as a function of $\bt_x^*$ for different cases; no crossing angle
  (Cr) and no hourglass (Hg), only the crossing angle, only the hourglass and with both effects. $\Phi = 0.5$ in
  all cases.   Bottom: Horizontal and vertical tune shifts as functions of
  of $\Phi$ and $\bt_x^*$.  $\bt_y^*$ is constant at 1mm in all figures. }
\label{fig: xixy}
\efig


The first plot in Fig.\ref{fig: Lumi} show the luminosity as a function of the so called Piwinski angle parameter
$\Phi =\tan( \theta_C/2)\sg_z/\sg_x^{*}$ for four values of $\bt_x^*$.
This shows that the luminosity is relatively flat upto $\Phi \sim 0.5$ which corresponds to $\theta_C = 21$
mrad or 69 times the beam divergence, a relatively large value. 
The second plot shows the luminosity as  function of  $\Phi$ and $\bt_x^*$ over the ranges $0 \le \Phi \le 5$ 
and $0.01 [m] \le \bt_x^* \le 0.20 [m]$ respectively.  This plot shows that the luminosity varies slowly as a function of $\Phi$ over
$0 \le \Phi \le 0.5$, more rapidly from $0.5 \le \Phi \le 2$ and then is relatively flat over $2 \le \Phi \le 5$ .
Decreasing $\bt_x^*$ from 0.2  m to 0.01 m increases the luminosity to nearly $4 \times 10^{34}$
cm$^{-2}$ s$^{-1}$ for $0 \le \Phi \le 0.5$. However, the vertical tune shift at these parameters is very large
at $\sim 0.25$, as the next figure shows.  

The top plot in Fig. \ref{fig: xixy} shows the vertical tune shift as a function of $\bt_x^*$ at constant $\Phi = 0.5$
for different cases showing the relative impact of the crossing angle and hourglass effects. It is clear that the
hourglass effect is dominant in determining the vertical tune shift. The bottom plots in this figure show the
horizontal and   vertical tune shifts as functions of $\bt_x^*$ and $\Phi$ with both effects included. The horizontal
tune shift $\xi_x$ varies more strongly with the crossing angle and is mostly independent of $\bt_x^*$. The
vertical tune shift on the other hand, varies strongly with $\bt_x^*$ and slowly with $\Phi$. 
 Assuming that tune shifts of $\sim 0.12$ are dynamically sustainable and the increased chromaticity can be
corrected, this suggests that
$\bt_x^* $ could be lowered to values in the range $0.025 \le \bt_x^* \le 0.05$m with $\bt_y^*= 0.001$ m.
These would increase luminosity to the range $(2 - 2.5) \times 10^{34}$ cm$^{-2}$s$^{-1}$.  
\bfig
\centering
\includegraphics[scale=0.5]{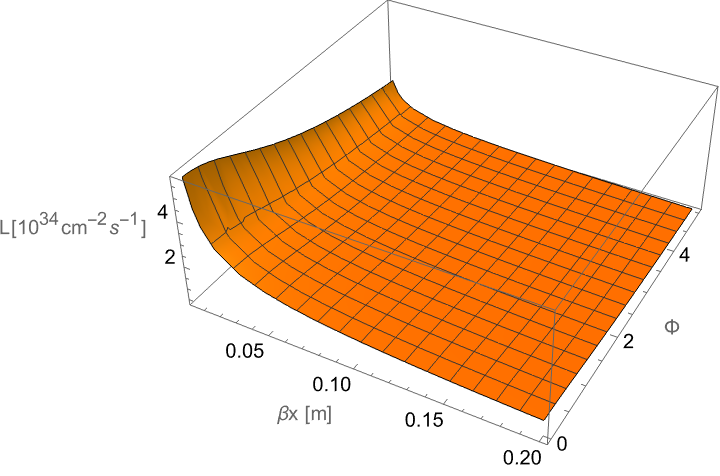}
\includegraphics[scale=0.5]{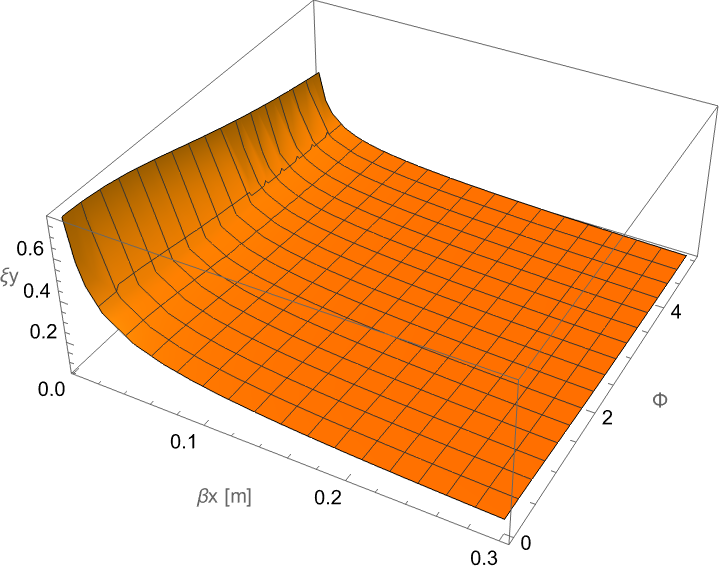}
\caption{Luminosity and $\xi_y$ as functions of $\bt_x^*$ and $\Phi$ at $\bt_y^*=0.5$mm.}
\label{fig: lumi_xiy_bety2}
\efig
We can be more aggressive by lowering $\bt_y^* $ further. The plots in Fig. \ref{fig: lumi_xiy_bety2} show that
with $\bt_x^* \le 0.01 \;{\rm m}, \; \bt_y^* = 0.0005$m, $\Phi < 2$, the vertical beam-beam tune shift $\xi_y \le 0.14$ and the luminosity increases to $\sim 4 \times 10^{34}$ cm$^{-2}$s$^{-1}$.   The major challenge
at these parameters will be to control the linear and non-linear IR chromaticities at these values of
$\bt_x^*, \bt_y^*$. 

\clearpage  

\section{FCC-ee collider}

We apply  the theory developed above to the FCC e$^+$-e$^-$ collider. The required parameters are shown in
Table \ref{table: FCC_param} taken from \cite{FCC_2019}.
\begin{table}[htb]
  \bec
  \btable{|c|c|} \hline
 Beam energy [GeV]  & 120 \\
  Circumference [km] &  97.75  \\
  Bunch intensity  &  1.8 $\times 10^{11}$ \\
  Number of bunches &  328  \\
  Emittance x [nm] / y [pm] & 0.63 / 1.3   \\
  $\bt_x^* / \bt_y^*$ [m]  & 0.3 / 0.001   \\
  $\sg_z$ [mm] &  5.3 \\
  Crossing angle (mrad) / Piwinski angle $\Phi$ & 30 / 5.8 \\
  \hline
  \etable
  \caption{Parameters of the FCC-ee}
  \label{table: FCC_param}
  \eec
  \bec
  \btable{|c|c|c|c|c||c|} \hline
  & No Cr or Hg & Hg only & Cr only & Hg and Cr &  FCC study \\  
  \hline
 ${\cal  L}$ [$10^{34}$ cm$^{-2}$s$^{-1}$] &  52 & 23  & 8.9  &  7.8   &  8.5 \\ 
  $\xi_x /\xi_y$ &  0.544 / 0.692  &  0.539 / 0.253 & 0.0158 / 0.118     & 0.0159 /  0.096   & 0.016 / 0.118 \\
    \hline
    \etable
 \caption{Luminosity and beam-beam tune shifts at the FCC parameter values used in the FCC study. }
    \label{table: FCC_lumibb}
 \eec
\end{table}
The bunch length in Table \ref{table: FCC_param} has been calculated with beamsstrahlung effects included,
according to reference \cite{FCC_2019}.

Table \ref{table: FCC_lumibb} shows the luminosity and beam-beam tuneshifts calculated with
different conditions in the several columns: (a) No crossing angle or hourglass, (b) hourglass only,
(c) crossing angle only, (d) both hourglass and crossing angle. These are compared with the values
found in the FCC study \cite{FCC_2019}. We find that our results with only the crossing angle are close to the
FCC study values but our values with both effects are lower, especially the vertical beam-beam tune shift which  is
more than 20\% lower. This is expected since $\bt_y^* \ll \sg_z$ while $\bt_x^* > \sg_z$. 

\bfig
\centering
\includegraphics[scale=0.45]{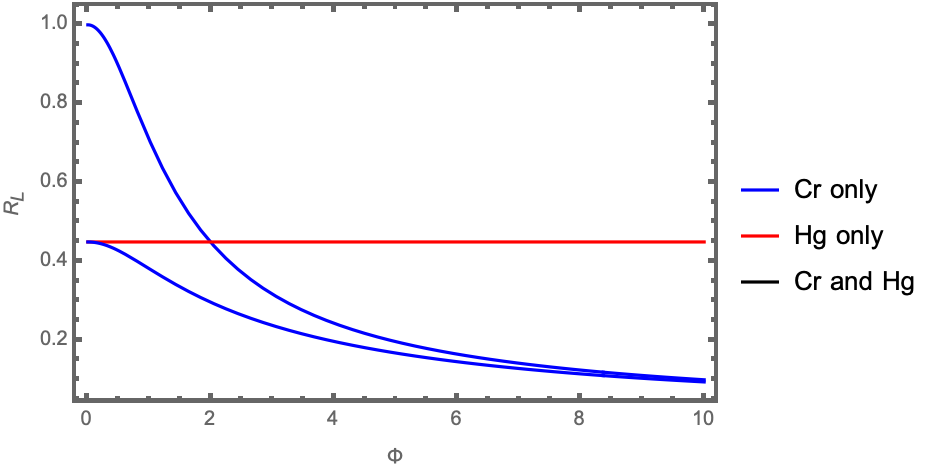}
\includegraphics[scale=0.45]{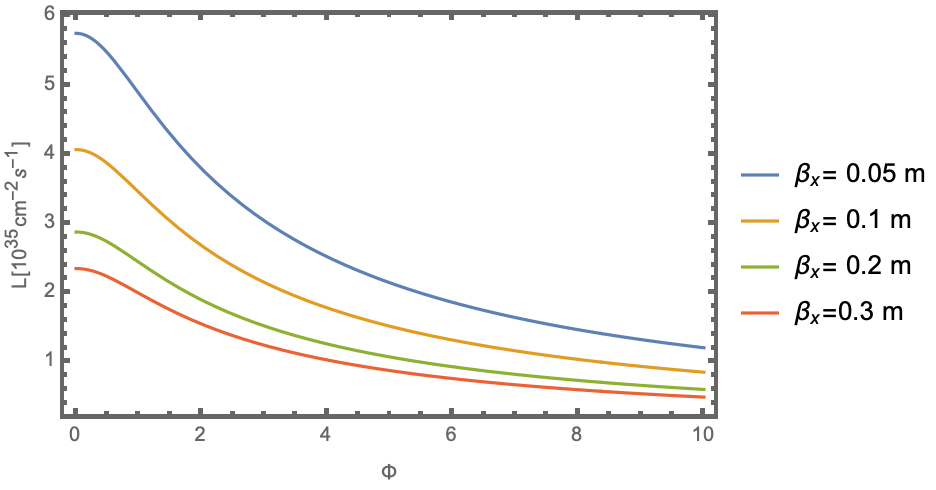}
\includegraphics[scale=0.8]{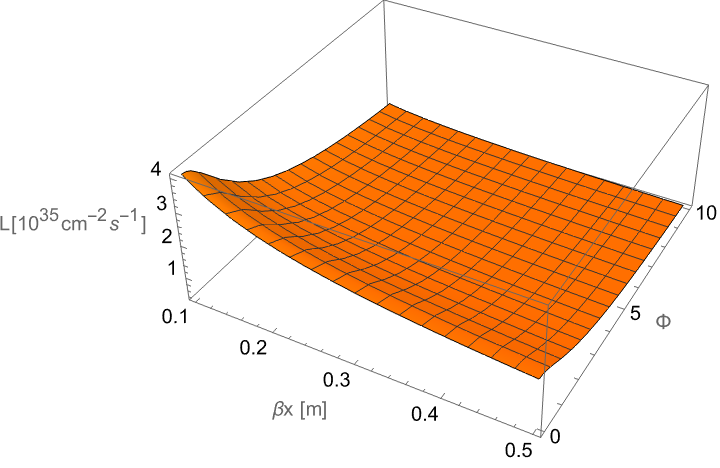}
\caption{Top Left: The correction factor $R_L$ vs  $\Phi$ for the cases (a) only the crossing angle (Cr), 
  (b) only the hourglass (hg), and (c) both crossing angle and the hourglass. Top right:  Luminosity as a
  function of $\Phi$ for different values of $\bt_x^*$ with both effects. Bottom:  Luminosity as a function
  of $\bt_x^*$ and $\Phi$. }
\label{fig: lumi_FCC} 
\efig
The left plot in Fig. \ref{fig: lumi_FCC} shows the luminosity correction factor $R_L$ from
Eq. (\ref{eq: RL_gen}) as a function of the Piwinski angle $\Phi$ at nominal values of $\bt_x^*, \bt_y^*$
for the cases with only the hourglass factor, only the crossing angle and with both effects included.
For the FCC parameters, the reduction due to the crossing angle exceeds the reduction due to the hourglass
factor when $\Phi > 2$. The right plot shows the luminosity as a function of $\Phi$ for different values of
$\bt_x^*$.  We observe, for example, that the luminosity at say $\bt_x^* = 0.05 $ m is in the range
$ 1.2 \le {\cal L}$ [ cm$^{-2}$ s$^{-1}$] $ \le 6 \times 10^{35 }$ while the range at the nominal $\bt_x^* = 0.3$m
is $ 0.49 \le {\cal L}$ [ cm$^{-2}$ s$^{-1}$] $ \le 2.5 \times 10^{35 }$ . The increase in luminosity at
lower $\bt_x^*$ decreases at larger crossing angles. 

\bfig
\centering
\includegraphics[scale=0.45]{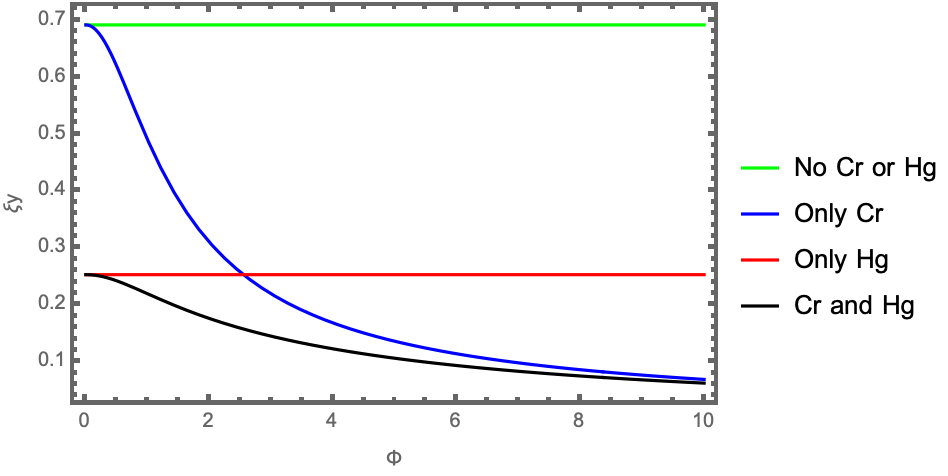}
\includegraphics[scale=0.46]{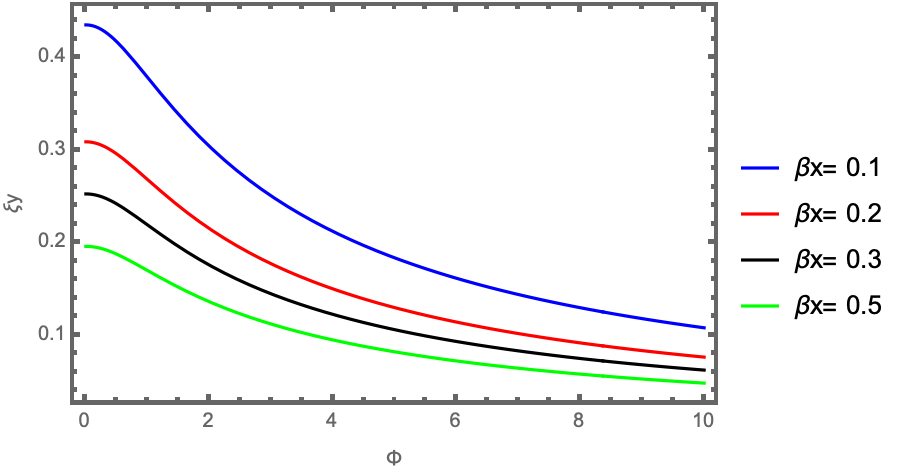}
\includegraphics[scale=0.5]{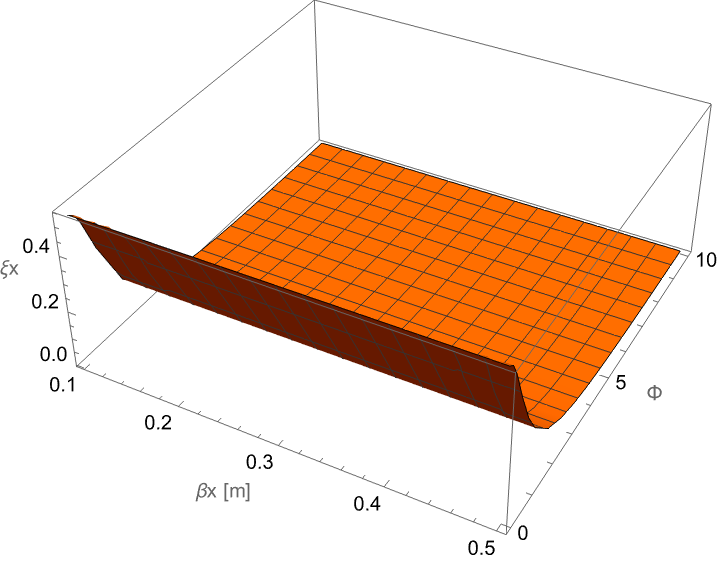}
\includegraphics[scale=0.35]{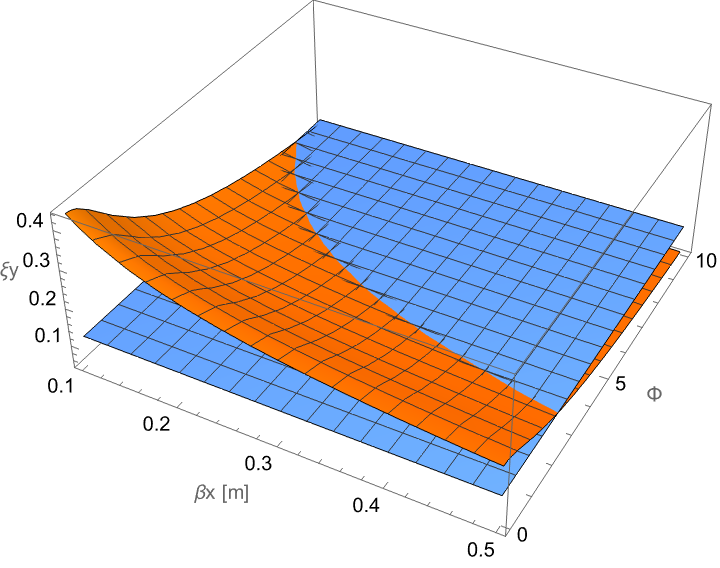}
\caption{Top row : Vertical beam-beam parameter as a function of $\Phi$ at $\bt_x^*=0.3$ m for
  different conditions (left) and
  different $\bt_x^*$ values (right). Bottom: $ \xi_x$ (left), and $\xi_y$
  (right) as functions of  $\bt_x^*$ and $\Phi$. In the right figure is also shown the plane (in blue) at
$\xi_y = 0.12$ intersecting the function $\xi_y(\Phi, \bt_x^*)$. }
\label{fig: xixy_FCC} 
\efig

\bfig
\centering
\includegraphics[scale=0.5]{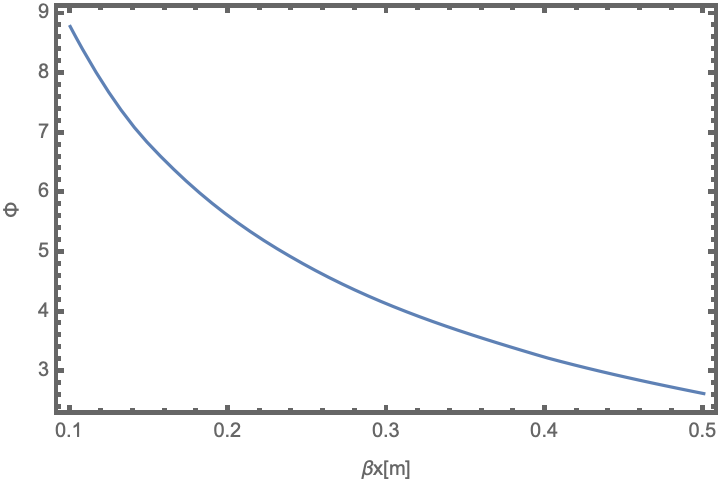}
\includegraphics[scale=0.35]{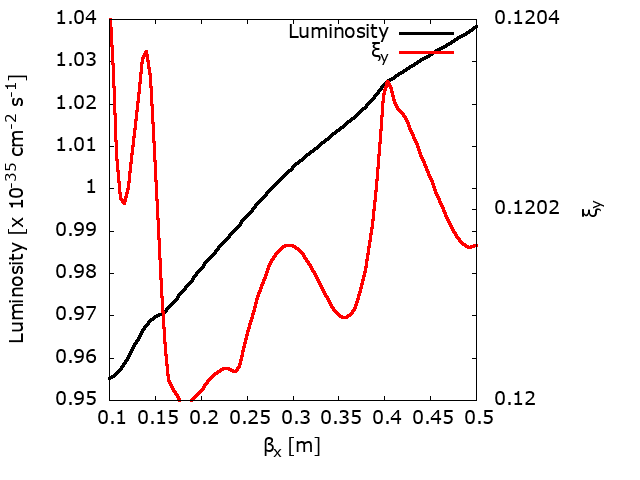} 			
\caption{Top: $\Phi$ as a function of $\bt_x$ at a constant value of $\xi_y = 0.12$.
  Bottom: Luminosity (left  vertical axis) and   $\xi_y$ (right vertical axis) as functions of $\bt_x^*$ with
the crossing angle determined   by the top figure.}
\label{fig: lumi_xiy_FCC} 
\efig
The top plots in Fig. \ref{fig: xixy_FCC} show $\xi_y$ as a function of $\Phi$ with or without the two effects
(left) and for different values of $\bt_x$.  The left plot shows that at $\bt)x^*= 0.3$ and at zero crossing
angle, $\xi_y = 0.67$ and 0.22 without and with the hourglass effect respectively. This value drops to 0.096
at the nominal crossing angle. The hourglass effect therefore significantly reduces the vertical beam-beam
tuneshift. The right plot shows,  among other observations, that $\xi_y$ is more  sensitive to the value of
$\bt_x^*$ at smaller $\Phi$ than at larger $\Phi$. The bottom plots in Fig. \ref{fig: xixy_FCC} show
$\xi_x, \xi_y$ as functions of $\bt_x^*, \Phi$. As expected, $\xi_x$ is insensitive to $\bt_x^*$ and falls
sharply with increasing $\Phi$. The right plot shows $\xi_y(\bt_x^*, \Phi)$ intersected with a planar surface
(in blue) at $\xi_y = 0.12$, assumed to be a target value.
The intersection of the the surfaces is a curve $\Phi$ as a function of $\bt_X^*$ along which $\xi_y = 0.12$.
The left plot in Fig. \ref{fig: lumi_xiy_FCC} shows this curve explicitly, calculated numerically. Assuming that
the bunch intensity is kept constant, this curve can be used to find the  values of $\bt_x^*, \Phi$ that
maximize the luminosity while keeping $\xi_y$ constant. The right plot in this figure shows the luminosity
(in black) is ${\cal L} \sim 10^{35}$ cm$^{-2}$ s$^{-1}$ over the range $0.1 \le \bt_x^* \le 0.5$ with the
corresponding $\Phi$ found from the left plot. Compared to the luminosity
${\cal L} = 0.79\times 10^{34}$ cm$^{-2}$ s$^{-1}$, this represents a 26\% increase in luminosity. This is
comparable to the 25\% luminosity  increase  achievable by increasing the bunch intensity to increase
$\xi_y$ from 0.096 to 0.12 at the nominal $\bt_x^*$ and $\Phi$. However the first method increases
$\bt_x^*$ from 0.3 to 0.5 and lowers the crossing angle from 30 mrad to $\sim 15$ mrad. This has additional
benefits of keeping the synchrotron radiation power constant, lowering the IR chromaticity and reducing aperture restrictions, and other  effects from
the smaller crossing angle and avoids potential problems at higher currents.


\section{Conclusions}

We developed exact expressions for the luminosity and beam-beam tuneshifts with both crossing angle and
hourglass effects.  We  showed that the expressions reduce to known expressions in the limit that one or the
other effect is absent. 

As mentioned earlier, the Fermi site filler design is very preliminary and detailed studies of the dynamics with
beam-beam interactions need to be done. The nominal design has one Interaction Point, a zero crossing angle
and achieves a peak luminosity $\sim 10^{34}$ cm$^{-2}$ s$^{-1}$. We find that with a crossing angle
around 21 mrad, $0.025 \le \bt_x^* \le 0.05$m would increase luminosity to the range
$(2 - 2.5) \times 10^{34}$ cm$^{-2}$s$^{-1}$ with $\xi_y$ nearly constant at 0.12. Since there is only one IP,
it is possible that $\xi_y$ could be increased from this value with an accompanying increase in the
luminosity. 

The FCC-ee design is considerably more advanced. Applying the theory developed in this report, we have the
following observations:
\bit
\item The crossing angle and the hourglass effect together reduce the luminosity and vertical beam-beam tuneshift compared to
  the values with the crossing angle alone. With both effects, the luminosity is $7.8 \times 10^{34}$
  cm$^{-2}$ s$^{-1}$  compared to $8.9 \times 10^{34}$ cm$^{-2}$ s$^{-1}$ with only the crossing angle.
  The values with only the crossing angle are close to those obtained in the FCC study. 
More significantly, the values of
  $\xi_y$ under the same conditions are 0.096 and 0.118 respectively.
\item Assuming a target value of $\xi_y = 0.12$, this suggests that the luminosity can be increased from
  the present value.
\item We find the luminosity can be increased by $\sim 25\%$ to $1 \times 10^{35}$ cm$^{-2}$ s$^{-1}$
  by simultaneously
  decreasing the crossing angle to $\theta_C \sim 15$ mrad and increasing $\bt_x^*$ to 0.5 m while keeping the
  vertical   tuneshift at 0.12. This would be better than the alternative method of increasing the
  bunch intensity while keeping the nominal values $\theta_C = 30$ mrad and  $\bt_x^* = 0.3$m .
  \eit

\noi   {\large \bf Acknowledgments} \\
Fermilab is operated by Fermi Research Alliance LLC under DOE contract No. DE-AC02CH11359.

\section{Appendix: Integration results}
\label{section: appendix_A}
\setcounter{equation}{0}
\renewcommand{\theequation}{A.\arabic{equation}}

The integrations for the tune shift without crossing angle or hourglass effects use
\beqr
\int_0^{1} du \; \left[ \fr{1}{(1 + (r_{yx}-1) u )} \right]^{1/2} & = & 
  2 \fr{ \sg_x^{*}}{\sg_y^{*} + \sg_x^{*}}  \label{eq: int_xix_nom} \\
  \int_0^{1} du   \left[\fr{1}{(1 + (r_{yx}-1)u)^3} \right]^{1/2}  & = & 
2 \fr{\sg_x^{*, 2}}{\sg_y^{*}(\sg_y^{*}+ \sg_x^{*})}   \label{eq: int_xiy_nom}
  \eeqr

The integrations for the tune shifts with crossing angle only use the results below
\beqr
\int_{0}^{\infty}  \fr{dt}{\sqrt{2\pi}} \exp[- a^2 t^2] (1 + b t^2) & = & \fr{1}{4\sqrt{2}}  \fr{2 a^2 + b}{a^3}   \\
 \int_0^1 \fr{dz}{\sqrt{(1 + U's)(1 + C z)^3}} & =  &  \fr{2}{(\sqrt{A+1} + \sqrt{C+1})\sqrt{1 + C}}
 \eeqr
Using these results,   the integrals over $t$ for $\xi_x, \xi_y$ yield
 \beqr
  \int_{0}^{\infty}\fr{dt}{\sqrt{2\pi}} \exp[-  (c^2 +  a^2 u ) t^2 /t_x^2 ]
  \left[ 1  + 2 S_C^2 ( 1  -  r_{sx} u ) t^2/t_x^2 \right]
  & = & \fr{1}{4\sqrt{2}} t_x \fr{2c^2 + 2  S_C^2 }{(c^2 + a^2 u)^{3/2}} \label{eq: xi_cr1}  \\
  \int_{0}^{\infty}\fr{dt}{\sqrt{2\pi}}  \exp[- c_c^2 t^2] \left[ 1 - 2c^2 t^2 \right]  & = &  0  \label{eq: xi_cr2}     \\
  \int_{0}^{\infty}\fr{dt}{\sqrt{2\pi}}  \exp[-  (c^2 + a^2 u) t^2] \left[ 1 - 2c^2 t^2 \right]
  & = &     \fr{1}{4\sqrt{2}}  \fr{2 a^2 u }{(c^2 + a^2 u)^{3/2}}  \label{eq: xi_cr3} 
  \eeqr

Beam-beam   tune shifts for flat beams
\beqr
  \int_0^{\infty} dt \;  \exp[- a^2 t^2] \fr{1 + b t^2 }{\sqrt{c + d t^2}} & = &  \fr{1}{4 a^2 \sqrt{d}}
  \left\{  2 a^2 \exp[a^2 c)/(2 d)] K_0( \fr{a^2 c}{2 d}) +  b \sqrt{\pi} U(\half, 0, \fr{a^2 c}{d}) \right\}
  \label{eq: flat_1} \non \\
  \mbox{}  \\ 
\int_0^{\infty} dt \;  \exp[- a^2 t^2] \fr{1 }{(c + d t^2)^{3/2}} & = &
\fr{\sqrt{\pi}}{2 c \sqrt{d}}U(1/2, 0, \fr{a^2 c}{d} )    \label{eq: flat_2} 
\eeqr
where $U(\half, 0, x)$ is the confluent geometric function which, to leading order, decays as $1/\sqrt{x}$.

\end{document}